\documentclass[aps,prd,preprint,superscriptaddress,showpacs]{revtex4}

\usepackage{graphicx}

\def\ld{\lambda}

\def\nnb{\nonumber}

\def\bwt{\begin{widetext}}
\def\ewt{\end{widetext}}
\def\be{\begin{equation}}
\def\ee{\end{equation}}
\def\bea{\begin{eqnarray}}
\def\eea{\end{eqnarray}}
\def\bean{\begin{eqnarray*}}
\def\eean{\end{eqnarray*}}
\def\bary{\begin{array}}
\def\eary{\end{array}}
\def\bit{\begin{itemize}}
\def\eit{\end{itemize}}

\def\ld{\lambda}

\def\su5u1{SU(5) \times U(1)}
\def\fsu5u1{SU(5) \times U(1)'}
\def\so10{SO(10)}
\def\sq20{SO(10) \times SO(10)}

\begin{document}

\title{Embedding Flipped $SU(5)$ into $SO(10)$}

\author{Chao-Shang Huang}

\affiliation{Institute of Theoretical Physics, Chinese Academy of Sciences,
 Beijing 100080, P. R. China}

\author{Tianjun Li}

\affiliation{Institute of Theoretical Physics, Chinese Academy of Sciences,
 Beijing 100080, P. R. China}

\affiliation{Department of Physics and Astronomy, Rutgers University, 
Piscataway, NJ 08854, USA}

\author{Chun Liu}

\affiliation{Institute of Theoretical Physics, Chinese Academy of Sciences,
 Beijing 100080, P. R. China}

\author{Jonathan P. Shock}

\affiliation{Institute of Theoretical Physics, Chinese Academy of Sciences,
 Beijing 100080, P. R. China}

\author{Feng Wu}

\affiliation{Institute of Theoretical Physics, Chinese Academy of Sciences,
 Beijing 100080, P. R. China}

\author{Yue-Liang Wu}

\affiliation{Institute of Theoretical Physics, Chinese Academy of Sciences,
 Beijing 100080, P. R. China}

\date{\today}

\begin{abstract}

We embed the flipped $SU(5)$ models into the 
$SO(10)$ models. After the $SO(10)$ gauge symmetry is
broken down to the flipped $SU(5)\times U(1)_X$ gauge symmetry,  
we can  split the five/one-plets and ten-plets in the spinor 
$\mathbf{16}$ and $\mathbf{\overline{16}}$ Higgs fields 
via the stable sliding singlet mechanism.
As in the flipped $SU(5)$ models,
 these ten-plet Higgs fields can break the flipped
$SU(5)$ gauge symmetry down to the Standard Model gauge
symmetry. The doublet-triplet splitting problem can be solved 
naturally by the missing partner mechanism, and 
 the Higgsino-exchange mediated proton decay can be suppressed elegantly.
Moreover, we show that there exists one pair of the light Higgs doublets
for the electroweak gauge symmetry breaking. 
Because there exist two pairs of additional vector-like 
particles with similar intermediate-scale masses,
the $SU(5)$ and $U(1)_X$ gauge couplings can be unified at the GUT scale 
which is reasonably (about one or two orders) 
higher than the $SU(2)_L\times SU(3)_C$ unification scale.
Furthermore, we briefly discuss the  simplest $SO(10)$ 
model with flipped $SU(5)$ embedding, and point
out that it can not work without fine-tuning.

\end{abstract}

\pacs{11.25.Mj, 12.10.Kt, 12.10.-g}

\preprint{ RUNHETC-06-03, hep-ph/0606087}

\maketitle

\section{Introduction}
The gauge hierarchy problem is one of the main motivations to study the
physics beyond the Standard Model (SM). The Higgs boson is needed in the SM to break the 
electroweak gauge symmetry and give masses to 
the SM fermions, and the breaking scale  is 
directly related to the Higgs boson mass. 
However, in quantum field theory, the fermionic masses can be 
protected against quantum corrections by chiral symmetry, 
while there is no such symmetry for bosonic masses. The Higgs boson
 mass (squared) has a quadratic divergence at one loop,
and it is unnatural to make a stable weak scale which is hierarchically 
smaller than the Planck scale.
Moreover, an aesthetic motivation for physics beyond the SM is 
Grand Unified Theories (GUTs) because GUTs can unify all the known
gauge interactions, and can give us a simple
understanding of the quantum numbers of the SM fermions, etc.

Supersymmetry  provides an elegant solution to the gauge
hierarchy problem. And the success of gauge coupling unification in
the Minimal Supersymmetric Standard Model (MSSM) strongly supports the
possibility of supersymmetric 
GUTs~\cite{Langacker:1990jh, Amaldi:1991cn}.  Other appealing features 
in supersymmetric GUTs are that
the electroweak gauge symmetry is broken by radiative corrections
due to the large top quark Yukawa coupling, and that the tiny neutrino
masses can be naturally generated by the see-saw mechanism~\cite{Seesaw}.
Therefore, supersymmetric GUTs are promising candidates that can
describe all the known fundamental interactions in nature except gravity.
However, there are severe problems in the four-dimensional 
supersymmetric GUTs, especially the
doublet-triplet splitting problem and the proton decay problem.

Among the known supersymmetric GUTs, 
only the flipped $SU(5)$ models
can naturally explain the doublet-triplet splitting via a simple 
and elegant missing partner mechanism~\cite{smbarr, dimitri, Huang:2003fv}. 
The Higgsino-exchange
mediated proton decay problem, which is such a difficulty for
the other supersymmetric GUTs, is solved automatically. However, the gauge 
group of flipped $SU(5)$ models is the product group $SU(5)
\times U(1)_X$, not a simple group, so the unifications
of the gauge interactions and their couplings are not ``grand". As a result, 
SM fermions in each family do not sit in a single
representation of the gauge group, unlike the case in the $SO(10)$ model. 
In flipped $SU(5)$ models, since the masses of
down-type quarks and charged leptons come from
different Yukawa couplings, the bottom quark mass
is generically not equal to the $\tau$ lepton mass at the GUT scale, 
which is one of the consistent predictions in
the other supersymmetric GUTs, e.g., $SU(5)$.
The grand unification of the gauge interactions,
and the unification of each family of the SM fermions into a single representation 
can be achieved by embedding the flipped $SU(5) $ into $SO(10)$. However, 
it is well-known that  
the missing partner mechanism can not work,
 because the partners that were missing in the $SU(5)\times U(1)_X$ 
multiplets are indeed appear in the larger $SO(10)$ multiplets.
To solve this problem, two kinds of models were proposed:
the five-dimensional orbifold $SO(10)$ models~\cite{Barr:2002fb}, and
the four-dimensional $SO(10)\times SO(10)$ models with
bi-spinor link Higgs fields~\cite{Huang:2004ui} (For
other $SO(10)$ models with flipped 
$SU(5) $ embedding, please 
see Refs.~\cite{Chamoun:2001in}.).

In this paper, we would like
to embed the flipped $SU(5)$ models into
the four-dimensional $SO(10)$ models where the missing
partner mechanism can still work elegantly.
In the flipped $SU(5)$ models,
the Higgs fields $H$ and $\overline{H}$, which break the
flipped $SU(5)$ gauge symmetry down to
the SM gauge symmetry, are one pair of vector-like
fields in the 
$(\mathbf{10}, \mathbf{1})$ and $(\mathbf{\overline{10}}, \mathbf{-1})$ 
 representations of $SU(5)\times U(1)_X$, respectively. When we 
embed the flipped $SU(5)$ into $SO(10)$, these Higgs fields 
 $H$ and $\overline{H}$ respectively are embedded 
into the Higgs fields 
$\Sigma$ and $\overline{\Sigma}$ in the spinor $\mathbf{16}$
and $\mathbf{\overline{16}}$ representations of $SO(10)$.
The missing partners for the MSSM Higgs doublets $H_u$ and $H_d$ 
respectively belong
to the $(\mathbf{\overline{5}}, \mathbf{-3})$ and 
$({\mathbf{5}}, \mathbf{3})$ of the $\Sigma$
and $\overline{\Sigma}$ when we decompose the 
$SO(10)$ spinor representations into the $SU(5)\times U(1)_X$
representations (for detail decompositions please see Appendix A).
Also, in the flipped $SU(5)$ models,
the Higgs fields $\overline{h}$ and $h$, which
include the Higgs doublets $H_u$ and $H_d$, are in 
$(\mathbf{\overline{5}}, \mathbf{2})$ and 
$(\mathbf{5}, \mathbf{-2})$ representations, respectively.
 Interestingly, the Higgs fields $\overline{h}$ and $h$ 
in our models can form a
 ${\mathbf{10}}$ representation Higgs field $h_{\mathbf{10}}$ of $SO(10)$. 
Note that we will break
the $SO(10)$ gauge symmetry down to the flipped $SU(5)$
gauge symmetry at the GUT scale $M_{GUT}$, and further down to the
SM gauge symmetry at the $SU(2)_L\times SU(3)_C$ unification
scale $M_{23}$. So, to have the 
 successful missing partner mechanism for the doublet-triplet splitting,
we must split the five-plets and ten-plets in the $\Sigma$
and $\overline{\Sigma}$, {\it i.~e.}, the five-plets
in the $\Sigma$ and $\overline{\Sigma}$ must have mass around the scale $M_{GUT}$
while the corresponding ten-plets should remain massless after
the $SO(10)$ gauge symmetry breaking.

We construct the three-family $SO(10)$ models with 
two adjoint Higgs fields $\Phi$ and $\Phi'$, 
$\Sigma$, $\overline{\Sigma}$, $h_{\mathbf{10}}$, one pair of 
spinor $\mathbf{16}$ and $\mathbf{\overline{16}}$ representations $\chi$
and $\overline{\chi}$, and several singlets.
After the $SO(10)$ gauge symmetry is broken down to the
flipped $SU(5)$ gauge symmetry,  
 the five/one-plets and ten-plets in the multiplets
$\overline{\chi}$ and $\Sigma$, and  $\overline{\Sigma}$ and $\chi$ 
can be splitted via 
 the sliding singlet mechanism. And we can show that this
sliding singlet mechanism is stable. Similar to the
flipped $SU(5)$ models, we can break the
gauge symmetry down to the SM gauge symmetry by giving
vacuum expectation values (VEVs) to the neutral 
singlet components of $H$ and $\overline{H}$.
The doublet-triplet splitting can be 
realized by the simple missing partner mechanism, and the 
Higgsino-exchange mediated proton decay is negligible.
Moreover, we show that there exists one pair of the light
Higgs doublets mainly from $H_u$ and $H_d$ for
the electroweak gauge symmetry breaking.
Since there exist two pairs of  vector-like particles 
(mainly from the correspoding components in
 $\chi$ and $\overline{\chi}$)
with roughly the same intermediate-scale masses
whose SM quantum numbers are 
$\left({\mathbf{(3, 2, {1\over 6}), ({\bar 3}, 2,
-{1\over 6})}} \right)$ and
$\left({\mathbf{({\bar 3},  1, {1\over 3}) + ({3},
1, -{1\over 3})}} \right)$,
the $SU(5)\times U(1)_X$ gauge coupling unification can be achieved at the GUT scale
which is  reasonably (about one or two orders) higher than 
the $SU(2)_L\times SU(3)_C$ unification scale~\cite{Lopez:1993qn, Lopez:1995cs}.
Therefore, we can keep the beautiful features and get rid of the drawbacks 
of the flipped $SU(5)$ models in our $SO(10)$ models.

Furthermore, we briefly consider the simplest $SO(10)$ model 
with flipped $SU(5)$ embedding, and point
out that we have to fine-tune some mass
parameters so that the model can be consistent. 
We also explain how to generate the suitable vector-like mass 
for $\chi$ and $\overline{\chi}$.

This paper is organized as follows: in Section II we briefly review
the flipped $SU(5)$ models, and the sliding
singlet mechanism. We present our $SO(10)$ models in Section III.
Moreover, we consider the mixings between the light and superheavy 
particles, and study gauge coupling unification in Section IV.
Our remarks on the simplest $SO(10)$ model and the vector-like mass 
for $\chi$ and $\overline{\chi}$
are given in Section V. Section VI is our discussion and
conclusions. We present the $SO(10)$ generators
in the spinor representations in Appendix A.

\section{Brief Review}

In this Section, we would like to briefly review the flipped
$SU(5)$ models~\cite{smbarr, dimitri}, 
and the sliding singlet mechanism~\cite{Witten:1981kv}.

\subsection{The flipped $SU(5)$ Models}

First, let us consider the flipped
$SU(5)$ models~\cite{smbarr, dimitri}. 
We can define the generator $U(1)_{Y'}$ in $SU(5)$ as 
\bea 
T_{\rm U(1)_{Y'}} \equiv {\rm diag} \left(-{1\over 3}, -{1\over 3}, -{1\over 3},
 {1\over 2},  {1\over 2} \right),
\label{u1yp}
\eea
and the hypercharge is given by
\bea
Q_{Y} = {1\over 5} \left( Q_{X}-Q_{Y'} \right).
\label{ycharge}
\eea

There are three families of the SM fermions with the following
$SU(5)\times U(1)_{X}$ quantum numbers
\bea
F_i={\mathbf{(10, 1)}},~ {\bar f}_i={\mathbf{(\bar 5, -3)}},~
{\bar l}_i={\mathbf{(1, 5)}},
\label{smfermions}
\eea
where $i=1, 2, 3$.
As an example, the particle assignments for the first family are
\bea
F_1~=~(Q_1, ~D^c_1, ~N^c_1)~,~~{\overline f}_1~=~(U^c_1, ~L_1)~,~
~{\overline l}_1~=~E^c_1~,~
\label{smparticles}
\eea
where $Q$ and $L$ are respectively the superfields of the left-handed
quark and lepton doublets, $U^c$, $D^c$, $E^c$ and $N^c$ are the
$CP$ conjugated superfields for the right-handed up-type quark,
down-type quark, lepton and neutrino, respectively. In addition, 
to give heavy masses to the right-handed neutrinos,
we add three singlets $\phi_i$.

To break the GUT and electroweak gauge symmetries, we introduce two pairs
of vector-like Higgs fields
\bea
H~=~{\mathbf{(10, 1)}}~,~~{\overline{H}}={\mathbf{({\overline{10}}, -1)}}~,~
~h={\mathbf{(5, -2)}}~,~~{\overline h}={\mathbf{({\bar {5}}, 2)}}~.~\,
\label{Higgse1}
\eea
We label the states in the $H$ multiplet by the same symbols as in
the $F$ multiplet, and for ${\overline H}$ we just add ``bar'' above the fields.
Explicitly, the Higgs particles are
\bea
H~=~(Q_H, ~D_H^c, ~N_H^c)~,~~
{\overline{H}}~=~ ({\overline{Q}}_{\overline{H}}, ~{\overline{D}}^c_{\overline{H}}, 
~{\overline {N}}^c_{\overline H})~,~\,
\label{Higgse2}
\eea
\bea
h~=~(D_h, ~D_h, ~D_h, ~H_d)~,~~
{\overline h}~=~({\overline {D}}_{\overline h}, ~{\overline {D}}_{\overline h},
{\overline {D}}_{\overline h}, ~H_u)~,~\,
\label{Higgse3}
\eea
where $H_d$ and $H_u$ are the two Higgs doublets in the MSSM.

To break the $SU(5)\times U(1)_{X}$ gauge symmetry down to the SM
gauge symmetry, we introduce the following superpotential 
\bea
{ W}=\lambda_1 H H h + \lambda_2 {\overline H} {\overline H} {\overline
h} + S ({\overline H} H-M_{H}^2)~,~\,
\label{spgut} 
\eea 
where $S$ is a singlet, and
$\lambda_1$ and $\lambda_2$ are Yukawa couplings. There is only
one F-flat and D-flat direction, which can always be rotated along
the $N^c_H$ and ${\overline {N}}^c_{\overline H}$ directions. So, we obtain that
$\langle N^c_H \rangle = \langle {\overline {N}}^c_{\overline H} \rangle =M_H$. 
In addition, the superfields $H$ and ${\overline H}$ are eaten and acquire 
large masses via the Higgs mechanism with supersymmetry, except for $D_H^c$ and 
${\overline {D}}^c_{\overline H}$. The superpotential terms $\lambda_1 H H h$ 
and
$\lambda_2 {\overline H} {\overline H} {\overline h}$ combine the $D_H^c$ and
${\overline {D}}^c_{\overline H}$ with the $D_h$ and ${\overline {D}}_{\overline h}$,
respectively, to form the massive eigenstates with masses
$2 \lambda_1 \langle N^c_H \rangle $ and 
$2 \lambda_2 \langle {\overline {N}}^c_{\overline H} \rangle $. 
Since there are no partners in $H$ and $\overline{H}$ for $H_u$ and $H_d$,  we
naturally obtain the doublet-triplet splitting due to the missing
partner mechanism. Because the triplets in $h$ and ${\overline h}$ only have
small mixing through the $\mu$ term, the Higgsino-exchange mediated
proton decay are negligible, {\it i.e.},
we do not have the dimension-5 proton decay problem.

The SM fermion masses are from the following superpotential
\bea 
{ W}_{\rm Yukawa} = {1\over 2} y_{ij}^{D}
F_i F_j h + y_{ij}^{U \nu} F_i  {\overline f}_j {\overline
h}+ y_{ij}^{E} {\overline l}_i  {\overline f}_j h + \mu h {\overline h}
+ y_{ij}^{N} \phi_i {\overline H} F_j~,~\,
\label{potgut}
\eea
where $y_{ij}^{D}$, $y_{ij}^{U \nu}$, $y_{ij}^{E}$ and $y_{ij}^{N}$
are Yukawa couplings, and $\mu$ is the bilinear Higgs mass term.

After the $SU(5)\times U(1)_X$ gauge symmetry is broken down to the SM gauge 
symmetry, the above superpotential gives 
\bea 
{ W_{SSM}}&=&
y_{ij}^{D} D^c_i Q_j H_d+ y_{ji}^{U \nu} U^c_i Q_j H_u
+ y_{ij}^{E} E^c_i L_j H_d+  y_{ij}^{U \nu} N^c_i L_j H_u \nnb \\
&& +  \mu H_d H_u+ y_{ij}^{N} 
\langle {\overline {N}}^c_{\overline H} \rangle \phi_i N^c_j
 + \cdots (\textrm{decoupled below $M_{GUT}$}). 
\label{poten1}
\eea

\subsection{Sliding Singlet Mechanism}

The sliding singlet mechanism was originally proposed in 
the supersymmetric $SU(5)$ model~\cite{Witten:1981kv}, 
where the Higgs superpotential is
\begin{eqnarray}
W = W(\Phi) + {\overline H}_{\mathbf{\bar 5}} 
\left( \Phi + S \right) H_{\mathbf{5}}~,~\,
\end{eqnarray}
where $\Phi$ is an $SU(5)$ adjoint Higgs field, 
$S$ is a SM singlet, and
${\overline H}_{\mathbf{\bar 5}} $ and $H_{\mathbf{5}}$
are the anti-fundamental and fundamental Higgs fields
which respectively contain one pair of Higgs doublets $H_d$ and
$H_u$.

With suitable superpotential $W(\Phi)$ for $\Phi$, 
one assumes that $\Phi$ obtains the following VEV
\begin{eqnarray}
\Phi &=& {\rm diag}\left(-{1\over 3}, -{1\over 3}, -{1\over 3},
{1\over 2}, {1\over 2} \right) V_{\Phi}~.~\,
\end{eqnarray}
Then, the $SU(5)$ gauge symmetry is broken down to
the SM gauge symmetry.

The F-flatness conditions 
for the F-terms of ${\overline H}_{\mathbf{\bar 5}}$ and $ H_{\mathbf{5}}$,
which is valid at a supersymmetric minimum,
give the following equations
\begin{eqnarray}
\left( \langle \Phi \rangle + \langle S \rangle \right)  
\langle H_{\mathbf{5}} \rangle =0 ~,~
\langle {\overline H}_{\mathbf{\bar 5}} \rangle 
\left( \langle \Phi \rangle + \langle S \rangle \right) =0 ~.~\,
\end{eqnarray}
To break the electroweak gauge symmetry, the Higgs doublets
$H_d$ and $H_u$ are supposed to obtain VEVs around the electroweak scale,
From F-flatness conditions $F_{H_d}=F_{H_u}=0$, we obtain 
\begin{eqnarray}
\langle S \rangle &=& - {1\over 2} V_{\Phi}~.~\,
\end{eqnarray}
Therefore, we have 
\begin{eqnarray}
 \langle \Phi \rangle + \langle S \rangle 
&=&  {\rm diag}\left(-{5\over 6}, -{5\over 6}, -{5\over 6},
 0, 0 \right) V_{\Phi}~.~\,
\end{eqnarray}
As a result, the color triplets in ${\overline H}_{\mathbf{\bar 5}}$
and $H_{\mathbf{5}}$ will obtain vector-like mass around $V_{\Phi}$, 
while the doublets will remain massless after the $SU(5)$ gauge
symmetry breaking. Because  the singlet slides  to cancel off 
the VEV of the adjoint Higgs field in the $SU(2)_L$
block, this mechanism is called the sliding singlet mechanism.

However, the sliding singlet mechanism for supersymmetric $SU(5)$ model 
breaks down due to the supersymmetry breaking~\cite{Polchinski:1982an}. 
The potential from
the F-terms of ${\overline H}_{\mathbf{\bar 5}}$ and $ H_{\mathbf{5}}$
only gives the electroweak-scale mass 
(${\sqrt {(\langle H_d^0 \rangle)^2 + (\langle H_u^0 \rangle)^2}}$) to $S$,
and the soft supersymmetry breaking gives $S$ mass around the
supersymmetry breaking scale $M_S$. However,
$S$ couples to the triplets in ${\overline H}_{\mathbf{\bar 5}}$
and $H_{\mathbf{5}}$ with masses around the GUT scale, so, the
one-loop tadpole graphs with the triplets running around the
loop induce the following two terms in the potential
in the low energy effective theory
that destroy the above doublet-triplet splitting 
\begin{eqnarray}
T_1 ~=~{\cal O} (m_g^2 M_{GUT}) S + ~{\rm H.C.}~,~~
T_2 ~=~{\cal O} (m_g M_{GUT}) F_S + ~{\rm H.C.}~,~\,
\end{eqnarray}
where $m_g$ is the gravitino  mass, which is usually around $M_S$.

The $T_1$ term will shift the VEV of S from its supersymmetric
minimum $-V_{\Phi}/2$ by the following amount
\begin{eqnarray}
\delta \langle S \rangle ~\sim~
{{{\cal O} (m^2_g M_{GUT})}\over {{\cal O}({M}_S^2) + 
 (\langle H_d^0 \rangle)^2 + (\langle H_u^0 \rangle)^2}} 
~\sim~ {\cal O} (M_{GUT})~,~\,
\end{eqnarray}
and then the doublets in  ${\overline H}_{\mathbf{\bar 5}}$
and $H_{\mathbf{5}}$ will obtain the vector-like mass
around the GUT scale.

In addition, after we integrate out the auxiliary field
$F_S$, the $T_2$ term gives the following term in
the potential
\begin{eqnarray}
V & \supset & | {\overline H}_{\mathbf{\bar 5}} H_{\mathbf{5}}
+ {\cal O} (m_g M_{GUT}) |^2 ~.~\,
\end{eqnarray}
Thus, the VEVs of $H_d^0$ 
and $H_u^0$ are around  the scale ${\sqrt {m_g M_{GUT}}}$,
which is inconsistent with the known value of 
${\sqrt {(\langle H_d^0 \rangle)^2 + (\langle H_u^0 \rangle)^2}}
\simeq 246.2~{\rm GeV}$.

In the gauge mediated supersymmetry breaking scenario,
the gravitino mass can be very light and below the keV scale. 
However, the sliding singlet mechanism still may not work~\cite{Chikira:1998bq}.

The sliding singlet mechanism can be successfully
applied to the rank five or higher GUT 
groups~\cite{Sen:1984aq, Barr:1997pt, Maekawa:2003ka},
 for example, the $SU(6)$ and $E_6$
models, etc. The point is that the corresponding Higgs
fields like the ${\overline H}_{\mathbf{\bar 5}}$
and $H_{\mathbf{5}}$ in the $SU(5)$ model can have 
 the very large or GUT-scale VEVs.
Let us briefly comment on the $SU(6)$ models.
To keep the F-flatness and have  one pair of 
light Higgs doublets, we need at least three pairs of
vector-like particles in 
the $SU(6)$ fundamental $\mathbf{6}$ and anti-fundamental
$\mathbf{\overline 6}$ representations. In the known model,
there are four pairs of such particles~\cite{Barr:1997pt}.

\section{$SO(10)$ Models}

We will construct the $SO(10)$ models where the gauge symmetry
is broken down to the flipped $SU(5)$ gauge symmetry
by giving VEVs to the adjoint Higgs fields, and
further down to the SM gauge symmetry by giving VEVs
to the $H$ and $\overline{H}$. We denote the SM
fermions as $\psi_i$ which form the spinor $\mathbf{16}$
representation. We introduce two adjoint $\mathbf{45}$ representation
Higgs fields $\Phi$ and $\Phi'$, one pair of the spinor $\mathbf{16}$
and $\mathbf{\overline{16}}$ representation 
Higgs fields $\Sigma$ and ${\overline \Sigma}$,
one $\mathbf{10}$ representation Higgs field $h_{\mathbf 10}$,
one pair of the spinor $\mathbf{16}$ and
$\mathbf{\overline{16}}$ representation vector-like particles $\chi$ and
${\overline \chi}$, and nine singlets $\phi_i$, $S$, $S'$,
$S_i$, and $S_{\Sigma}$ where $i=1, 2, 3$.
The complete particle content is given 
in Table \ref{Spectrum-SO(10)}.

\renewcommand{\arraystretch}{1.4}
\begin{table}[t]
\caption{Particle content in $SO(10)$ models.
\label{Spectrum-SO(10)}}
\vspace{0.4cm}
\begin{center}
\begin{tabular}{|c|c|}
\hline 
Representation & Chiral Superfields  \\
\hline\hline
$\mathbf{45}$ & $\Phi$; ~~$\Phi'$ \\
\hline
$\mathbf{16}$ & $\psi_i$; ~~$\Sigma$; ~~$\chi$ \\
\hline
$\mathbf{\overline{16}}$ & ${\overline \Sigma}$;
~~${\overline \chi}$ \\
\hline
$\mathbf{10}$ & $h_{\mathbf 10}$ \\
\hline
$\mathbf{1}$ & ~~$\phi_i$; ~~$S$; ~~$S'$;
~~$S_i$; ~~$S_{\Sigma}$~~ \\
\hline 
\end{tabular}
\end{center}
\end{table}

In terms of the particles in the flipped $SU(5)$ models,
we have
\begin{eqnarray}
\psi_i~=~(F_i, ~{\bar f}_i, ~{\bar l}_i)~;~~
h_{\mathbf 10}~=~(h,~{\overline h})~.~\,
\end{eqnarray}

In our convention, for one pairs of the spinor $\mathbf{16}$
and $\mathbf{\overline{16}}$ representation chiral superfields
 $K$ and ${\overline K}$, we denote their components like 
the SM fermions as following
\begin{eqnarray}
K~=~(K_F, ~K_{\bar f}, ~K_{\bar l})~,~~
{\overline K} ~=~({\overline K}_{\overline F}, 
~{\overline K} _{f}, ~{\overline K} _{l})~,~\,
\end{eqnarray}
where
\begin{eqnarray} 
&& K_F~=~(Q_K, ~D_K^c, ~N_K^c)~,~~K_{\bar f}=(U^c_K, ~L_K)~,~ \nonumber \\ 
&& {\overline K}_{\overline F} ~=~({\overline{Q}}_{\overline{K}}, 
{\overline{D}}^c_{\overline{K}}, {\overline {N}}^c_{\overline K})
~,~~ {\overline K}_{f}=(\overline{U}^c_{\overline K}, 
~\overline{L}_{\overline K})~.~\,
\end{eqnarray}

The only exception is that similar to the flipped $SU(5)$ models,
 we denote the Higgs fields $\Sigma_F$ and 
${\overline \Sigma}_{\overline F}$ as $H$ and $\overline{H}$, respectively.
To be concrete, we have
\begin{eqnarray} 
\Sigma~=~(H, ~\Sigma_{\bar f}, ~\Sigma_{\bar l})~,~~
{\overline \Sigma} ~=~({\overline H}, 
~{\overline \Sigma} _{f}, ~{\overline \Sigma} _{l})~.~\,
\end{eqnarray}

The superpotential is 
\begin{eqnarray}
W &=& W(\Phi, \Phi') + W(\Sigma, {\overline \Sigma}) + 
y_{ij} \psi_i h_{\mathbf 10} \psi_j + y_{ij}^N \phi  {\overline \Sigma} \psi_j 
+{1\over 2} \mu h_{\mathbf 10} h_{\mathbf 10} 
+ \lambda_1 \Sigma h_{\mathbf 10} \Sigma
\nonumber \\ &&
+ \lambda_2 {\overline \Sigma} h_{\mathbf 10} {\overline \Sigma} 
+\lambda_3 {\overline \chi} (\Phi + \lambda_4 S) \Sigma 
+ \lambda_5 \overline{\Sigma} 
(\Phi' + \lambda_6 S') \chi + M_{\chi} {\overline \chi}
\chi~,~\,
\label{SO(10)-WSS}
\end{eqnarray}
where $y_{ij}$, $y_{ij}^N$, and $\lambda_{i}$ ($i=1, 2,..., 6$) are Yukawa couplings,
and $\mu$ and $M_{\chi}$ are vector-like masses. The general superpotential 
$W(\Phi, \Phi')$ for $\Phi$ and $\Phi'$, and the simple superpotential 
$W(\Sigma, {\overline \Sigma})$ for $\Sigma$ and ${\overline \Sigma}$ are
\begin{eqnarray}
W(\Phi, \Phi') &=& \kappa \Phi^3 + M_{\Phi} \Phi^2 
+ \lambda_7 S_1 (\Phi^2-m_{11}^2) 
+ \kappa' \Phi^{\prime 3} + M_{\Phi'} \Phi^{\prime 2} 
+ \lambda'_7 S_2 (\Phi^{\prime 2}-m_{22}^2) \nonumber \\ 
&&
+M_{\Phi \Phi^{\prime}} \Phi \Phi^{\prime} 
+\lambda_8 S_3 (\Phi \Phi'-m_{12}^2)~,~
\end{eqnarray}
\begin{eqnarray}
W(\Sigma, {\overline \Sigma}) &=& S_{\Sigma}^2 
(\Sigma {\overline \Sigma} -M_{H}^2)~,~
\end{eqnarray}
where $\kappa$, $\kappa'$, 
$\lambda_7$, $\lambda'_7$, and $\lambda_8$ are Yukawa couplings,
and $M_{\Phi}$, $M_{\Phi'}$, $M_{\Phi \Phi^{\prime}}$,
$m_{11}$, $m_{22}$, $m_{12}$, and
$M_{H}$ are mass parameters.

Let us briefly comment on $W(\Phi, \Phi')$. First, we must have
at least one term which couples $\Phi$ and $\Phi'$ so that
we only have one global $SO(10)$ symmetry in $W(\Phi, \Phi')$,
{\it i.~e.}, the $SO(10)$ gauge symmetry. Otherwise, we will have some
unwanted massless Nambu-Goldstone bosons. Second, some of the Yukawa couplings 
and mass parameters in $W(\Phi, \Phi')$ should be zero. For example,
$m_{11}$, $m_{22}$, and $m_{12}$ can not be all non-zero in general,
otherwise, we need to fine-tune these masses to satisfy the
F-flatness conditions $F_{S_i}=0$.
 Let us present a simple  $W(\Phi, \Phi')$
 \begin{eqnarray}
W (\Phi, \Phi' ) =  M_{\Phi \Phi'} \Phi \Phi'
 + \lambda_8 S_3 (\Phi \Phi' - m_{12}^2)~.~\,
\end{eqnarray}
The flatness of F-term of $S_3$ ($F_{S_3}=0$) implies that $\langle \Phi \rangle \not=0$ and $\langle \Phi' \rangle \not=0$. Also, the F-flatnesses of the F-terms of $\Phi$ and $\Phi'$ ($F_{\Phi} = F_{\Phi'} = 0$) imply that $\langle \Phi \rangle = \langle \Phi' \rangle \not=0$ and $\langle S_3 \rangle \not= 0$. By the way, at very high temperature, the 
$SO(10)$ gauge symmetry will be restored when we consider 
the superpotential at finite temperature.

The gauge fields of $SO(10)$ are in the adjoint representation 
of $SO(10)$ with dimension {\bf 45}. Under the gauge group
$SU(5)\times U(1)_X$, the $SO(10)$ gauge fields
decompose as~\cite{Group}
\begin{eqnarray}
{\bf 45= (24, 0) \oplus (10, -4) \oplus (\overline{10}, 4) \oplus
(1, 0)}~.~\,
\label{SO(10)-Decompose}
\end{eqnarray} 
To break the $SO(10)$ gauge symmetry down to the 
flipped $SU(5)$ gauge symmetry via
adjoint Higgs fields, we need to
give the VEVs to their singlet components.

As we explained in the Introduction, 
to achieve the doublet-triplet splitting via the missing
partner mechanism, we must split the
five/one-plets and ten-plets in the $\Sigma$ and $\overline{\Sigma}$
during the $SO(10)$ gauge symmetry breaking.
In order to give the GUT-scale masses to
 the $\Sigma_{\bar f}$, $\Sigma_{\bar l}$, 
${\overline \Sigma} _{f}$ and ${\overline \Sigma} _{l}$ while
keep $H$ and ${\overline H}$ massless when we break the $SO(10)$ gauge
symmetry  down to the flipped $SU(5)$ 
gauge symmetry, we should express the $SO(10)$ generators in 
the spinor representations which are $16\times 16$ matries and are 
given in Appendix A. 
Note that when the $U(1)_X$ generator $T_{U(1)_X}$ acts on the spinor 
representation $\mathbf{16}$, it gives us the corresponding $U(1)_X$ charges
of the particles belong to $\mathbf{16}$. So, we obtain
the generator for $U(1)_X$
\begin{eqnarray}
T_{U(1)_X} &=& {\rm diag}(1,1,1,-3,1,1,1,-3,1,1,1,5,-3,-3,-3,1)~.~\,
\end{eqnarray}

For simplicity, we  assume that
the $\Phi$ and $\Phi'$ obtain the VEVs at the GUT scale
due to the superpotential $W(\Phi, \Phi')$, and the F-flatness conditions for
the F-terms of
$\Phi$, $\Phi'$ and $S_i$ are satisfied by choosing suitable
Yukawa couplings and mass parameters in $W(\Phi, \Phi')$.
The explicit VEVs for $\Phi$ and $\Phi'$ are
\begin{eqnarray}
\langle \Phi \rangle &=& {\rm diag}(1,1,1,-3,1,1,1,-3,1,1,1,5,-3,-3,-3,1) 
~V_{\Phi}~,~\nonumber \\ 
\langle \Phi' \rangle &=& {\rm diag}(1,1,1,-3,1,1,1,-3,1,1,1,5,-3,-3,-3,1) 
~V_{\Phi'}~,~\,
\end{eqnarray} 
where $V_{\Phi}$ and $V_{\Phi'}$ are around the GUT scale.

The F-flatness conditions 
for the F-terms of ${\overline \chi}$ and $\chi$,
which is valid at a supersymmetric minimum,
give the following equations
\begin{eqnarray}
\left( \langle \Phi \rangle + \lambda_4 \langle S \rangle \right)  
\langle \Sigma \rangle =0 ~,~
\langle {\overline \Sigma} \rangle 
\left( \langle \Phi' \rangle + \lambda_6 \langle S' \rangle \right) =0 ~.~\,
\end{eqnarray}
To break the flipped $SU(5)$ gauge symmetry down to the SM
gauge symmetry, we give VEVs to 
 $N^c_H \subset H \subset \Sigma$ and 
${\overline {N}}^c_{\overline H} \subset {\overline H} \subset  {\overline \Sigma}$
 at the $SU(3)_C\times SU(2)_L$ unification scale $M_{23}$, which
is around $3.7\times 10^{16}$ GeV. 
From the F-flatness conditions $F_{N^c_H}=F_{{\overline {N}}^c_{\overline H}}=0$,
 we obtain
\begin{eqnarray}
\langle S \rangle ~=~ - {{V_{\Phi}} \over {\lambda_4}} ~,~
\langle S' \rangle ~=~ - {{V_{\Phi'}} \over {\lambda_6}} ~.~\,
\end{eqnarray}
Thus, we have
\begin{eqnarray}
\langle \Phi \rangle +  \lambda_4 \langle S \rangle 
&=& {\rm diag}(0,0,0,-4,0,0,0,-4,0,0,0,4,-4,-4,-4,0) ~V_{\Phi}~,~\nonumber \\ 
\langle \Phi' \rangle +  \lambda_6 \langle S' \rangle
&=& {\rm diag} (0,0,0,-4,0,0,0,-4,0,0,0,4,-4,-4,-4,0) ~V_{\Phi'}~.~\,
\end{eqnarray} 
Then we have the following vector-like
mass terms for the pairs
($\overline{\chi}_{f}$, $\Sigma_{\overline{f}}$),
($\overline{\chi}_{l}$, $\Sigma_{\overline{l}}$),
($\overline{\Sigma}_{f}$, $\chi_{\overline{f}}$), and
($\overline{\Sigma}_{l}$, $\chi_{\overline{l}}$)
\begin{eqnarray}
V & \supset & -4 \lambda_3 V_{\Phi} 
\left(\overline{\chi}_{f} \Sigma_{\overline{f}}
-\overline{\chi}_{l} \Sigma_{\overline{l}}\right)
-4 \lambda_5 V_{\Phi'}
\left(\overline{\Sigma}_{f} \chi_{\overline{f}}
- \overline{\Sigma}_{l} \chi_{\overline{l}} \right)~,~\,
\end{eqnarray} 
where for simplicity we neglect the $M_{\chi}$, which
will be shown to be very small compared to the scales 
$M_{GUT}$ and $M_{23}$ so that we can have one pair of
the light Higgs doublets for the electroweak gauge
symmetry breaking. However, 
the particles ${\overline \chi}_{\overline{F}}$,
$H$, ${\overline H}$ and $\chi_F$ are massless
if we neglect $M_{\chi}$.
Thus, we split the five/one-plets and ten-plets in 
the multiplets $\overline{\chi}$ and $\Sigma$, and 
 $\overline{\Sigma}$ and $\chi$ via the sliding singlet mechanism
after we break the $SO(10)$ gauge symmetry
down to the flipped $SU(5)$ gauge symmetry.

As discussed in the brief review of the flipped $SU(5)$ models,
we break the $SU(5)\times U(1)_{X}$ gauge symmetry down to the SM
gauge symmetry by giving VEVs to the
 $N^c_H$ and ${\overline {N}}^c_{\overline H}$ of $H$ and $\overline{H}$.
The superfields $H$ and ${\overline H}$ are eaten and acquire 
large masses via the Higgs mechanism with supersymmetry, except for $D_H^c$ 
and ${\overline {D}}^c_{\overline H}$. And the superpotential 
$\lambda_1 H H h \subset \lambda_1 \Sigma h_{\mathbf 10} \Sigma$ and
$\lambda_2 {\overline H} {\overline H} {\overline h}
\subset \lambda_2 {\overline \Sigma} h_{\mathbf 10} {\overline \Sigma} $
 combine the $D_H^c$ and
${\overline {D}}^c_{\overline H}$ with the $D_h$ and ${\overline {D}}_{\overline h}$,
respectively, to form the massive eigenstates with masses
$2 \lambda_1 \langle N^c_H \rangle $ and 
$2 \lambda_2 \langle {\overline {N}}^c_{\overline H} \rangle $. So, we
solve the doublet-triplet splitting problem naturally via the missing
partner mechanism. Because the triplets in $h$ and ${\overline h}$
of $h_{\mathbf 10}$ only have
small mixing through the $\mu$ term, the Higgsino-exchange mediated
proton decay are negligible, {\it i.~e.},
we do not have the dimension-5 proton decay problem.

Let us show that our sliding singlet mechanism is stable.
The $T_1$ type tadpoles will shift 
the VEVs of $S$ and $S'$ from its supersymmetric
minimum by the following amount
\begin{eqnarray}
\delta \langle S \rangle ~\sim~
{{{\cal O} (m^2_g M_{GUT})}\over 
{\lambda_3^2 \lambda_4^2 (\langle N^c_H \rangle)^2}}~,~
\delta \langle S' \rangle ~\sim~
{{{\cal O} (m^2_g M_{GUT})}\over {\lambda_5^2 \lambda_6^2
(\langle {\overline {N}}^c_{\overline H} \rangle)^2}}~.~\,
\end{eqnarray}
It is obvious that these shifting effects are tiny and
can be neglected.

Moreover, after we integrate out the auxiliary fields
$F_S$ and $F_{S'}$, the $T_2$ type tadpoles will give
us the following terms in the potential 
\begin{eqnarray}
V & \supset & |\lambda_3 \lambda_4 {\overline \chi} \Sigma + {\cal O} (m_g M_{GUT}) |^2 
+ |\lambda_5 \lambda_6 \overline{\Sigma} \chi + {\cal O} (m_g M_{GUT}) |^2~.~\,
\end{eqnarray}
Then, we obtain 
\begin{eqnarray}
\langle \overline{N}_{\overline{\chi}}^c \rangle 
~\sim~ - {{{\cal O} (m_g M_{GUT})}\over 
{\lambda_3 \lambda_4 \langle N_H^c \rangle}}~,~
\langle N_{\chi}^c \rangle ~\sim~ 
- {{{\cal O} (m_g M_{GUT})}\over 
{\lambda_5 \lambda_6 \langle \overline{N}_{\overline{H}}^c \rangle}}~.~\,
\end{eqnarray}
Because $\Sigma$ and $\overline{\Sigma}$, or $\chi$ and
$\overline{\chi}$ do not contain the one pair of
Higgs doublets $H_d$ and $H_u$ in the MSSM, it is fine that
we have very small  non-zero VEVs for $N_{\chi}^c$
and $ \overline{N}_{\overline{\chi}}^c $ 
compared to the scales $M_{GUT}$ and $M_{23}$.

Moreover, from the F-flatness conditions 
for the F-terms of ${\overline \chi}$ and $\chi$, we obtain
\begin{eqnarray}
 \langle \Phi \rangle + \lambda_4 \langle S \rangle ~\sim~
{{{\cal O} (m_g M_{GUT})}\over 
{\lambda_3 \lambda_5 \lambda_6 \langle \overline{N}_{\overline{H}}^c \rangle
\langle N_H^{c} \rangle}} ~M_{\chi}~,~~
 \langle \Phi' \rangle + \lambda_6 \langle S' \rangle  ~\sim~
{{{\cal O} (m_g M_{GUT})}\over 
{\lambda_3 \lambda_4 \lambda_5 \langle \overline{N}_{\overline{H}}^c \rangle
\langle N_H^{c} \rangle}}~M_{\chi}~.~\,
\end{eqnarray}
So, the variations on $\langle \Phi \rangle + \lambda_4 \langle S \rangle$ 
 and $\langle \Phi' \rangle + \lambda_6 \langle S' \rangle$ are also very small
compared to the scales $M_{GUT}$ and $M_{23}$, and will not affect the
splittings of the five/one-plets and ten-plets in the multiplets
$\overline{\chi}$ and $\Sigma$, and $\overline{\Sigma}$ and $\chi$.
Especially, for the gauge mediated
supersymmetry breaking, the gravitino mass can be around the keV scale,
and these variations are completely negligible. 
Therefore,  our sliding singlet mechanism is stable.
By the way, the VEVs of $\Phi$, $S$, $\Phi'$, $S'$, $N^c_H$,
 and ${\overline {N}}^c_{\overline H}$ will be shifted by tiny
amount due to non-zero $\langle N_{\chi}^c \rangle$ and
$\langle \overline{N}_{\overline{\chi}}^c \rangle$.

In the following discussions, for simplicity we will neglect
the VEVs of $N_{\chi}^c$ and $\overline{N}^c_{\overline{\chi}}$
that are very small compared to the $V_{\Phi}$, $V_{\Phi'}$,
$\langle N^c_H \rangle$, and 
$\langle {\overline {N}}^c_{\overline H} \rangle$.

\section{Phenomenological Consequences}

In this Section, we will study the mixings between the light and superheavy 
particles, and the gauge coupling unification.

\subsection{ Light and Superheavy Particle Mixings}

After the flipped $SU(5)$ gauge
symmetry breaking, the possible light particles are three families of
the SM fermions, one pair of the Higgs doublets $H_d$ and $H_u$, and 
one pair of the $\mathbf{10}$ representation $\chi_{F}$ 
and $\mathbf{\overline{10}}$ representation $\overline{\chi}_{\overline F}$
in $\chi$ and $\overline{\chi}$.
However, to make sure that $H_d$, $H_u$, $\chi_{F}$,
and $\overline{\chi}_{\overline F}$ are indeed light, 
we must calculate all the possible mixing mass matrices between
these particles and superheavy particles.
There are three types of relevant particle mixings:

(1) In the $SU(5)$ language, the doublets $(X, Y)$- and 
$(\overline{X}, \overline{Y})$-type particles in 
the $(\mathbf{24, 0})$ decomposed representations of the $\Phi$ and $\Phi'$ have 
the same SM quantum numbers as the quark doublet and its Hermitian conjugate.
After $N_{H}^c$ and $\overline{N}_{\overline{H}}^c$ obtain
VEVs, they will mix with the $Q_{\chi}$ and $\overline{Q}_{\overline{\chi}}$
in $\chi_F$ and $\overline{\chi}_{\overline F}$.
Let us denote the $(X, Y)$- and $(\overline{X}, \overline{Y})$-type particles 
in $\Phi$ as $Q_{\Phi}$ and $\overline{Q}_{\Phi}$, 
and in $\Phi'$ as $Q_{\Phi'}$ and $\overline{Q}_{\Phi'}$.
The mass terms in the superpotential are
\begin{eqnarray}
W & \supset & M_{XY}^{11} \overline{Q}_{\Phi} Q_{\Phi} 
+ M_{XY}^{12} \overline{Q}_{\Phi} Q_{\Phi'}
+  M_{XY}^{21} \overline{Q}_{\Phi'} Q_{\Phi} + M_{XY}^{22} \overline{Q}_{\Phi'} Q_{\Phi'} 
\nonumber \\ &&
+ \lambda_3 \langle N_{H}^c \rangle {\overline Q}_{\overline{\chi}} Q_{\Phi}
+ \lambda_5 \langle \overline{N}_{\overline H}^c \rangle \overline{Q}_{\Phi'} Q_{\chi}
+ M_{\chi} {\overline Q}_{\overline{\chi}} Q_{\chi} ~,~\,
\end{eqnarray}
where $M_{XY}^{ij}$ are the mass parameters around the GUT scale.
The corresponding mass matrix for the basis 
$(\overline{Q}_{\Phi}, \overline{Q}_{\Phi'}, {\overline Q}_{\overline{\chi}})^t$ versus
$(Q_{\Phi},  Q_{\Phi'}, Q_{\chi})$, where $t$ is transpose, are
the following
\begin{equation}
\label{MXYQ}
M_{XYQ\overline{Q}} ~=~\left(
\begin{array}{ccc}
M_{XY}^{11} & ~~M_{XY}^{12} & ~~0 \\
M_{XY}^{21} & ~~M_{XY}^{22} & 
~~\lambda_5 \langle \overline{N}_{\overline H}^c \rangle \\
\lambda_3 \langle N_{H}^c \rangle & ~~0 & ~~M_{\chi} \end{array}
\right)~.~\,
\end{equation}
The determinant of above mass matrix is
\begin{eqnarray}
{\rm Det}[M_{XYQ\overline{Q}}] &=& \left(M_{XY}^{11} M_{XY}^{22}
- M_{XY}^{12} M_{XY}^{21} \right) M_{\chi} \sim M_{GUT}^2 M_{\chi}~,~\,
\end{eqnarray}
where we assume that there is no fine-tuning.
So, there are two pairs of vector-like particles 
(major components belong to $Q_{\Phi}$ and $Q_{\Phi'}$, and
 $\overline{Q}_{\Phi}$ and $\overline{Q}_{\Phi'}$) with vector-like 
masses around the GUT scale, and one pair of 
vector-like particles (major components belong to $Q_{\chi}$ and 
${\overline Q}_{\overline{\chi}}$) with vector-like mass around $M_{\chi}$.

(2) The SM singlet mixings. For $\Phi$ and $\Phi'$, we consider
the $SU(5)\times U(1)_X$ singlets as given in Eq. (\ref{SO(10)-Decompose}),
corresponding to $U(1)_X$ gauge field component.
We denote the singlets in $\Phi$ and $\Phi'$ as $S_{\Phi}$ and
$S_{\Phi'}$. After the flipped $SU(5)$ gauge symmetry
breaking, we have the following mass terms in the superpotential for the
SM singlets $S_{\Phi}$, $S$, $S_{\Phi'}$, $S'$, $N_{\chi}^c$, and
$\overline{N}_{\overline{\chi}}^c$
\begin{eqnarray}
W & \supset & {1\over 2} M_{SX}^{11} S_{\Phi}^2 +  M_{SX}^{12} S_{\Phi} S_{\Phi'}
+ {1\over 2} M_{SX}^{22} S_{\Phi'}^2 
+  \lambda_3 \langle N_{H}^c \rangle  \overline{N}_{\overline{\chi}}^c 
(S_{\Phi} + \lambda_4 S)
\nonumber \\ &&
+  \lambda_5 \langle \overline{N}_{\overline H}^c  \rangle 
(S_{\Phi'} + \lambda_6 S') {N}_{{\chi}}^c 
+  M_{\chi} \overline{N}_{\overline{\chi}}^c {N}_{{\chi}}^c
~,~\,
\end{eqnarray}
where $M_{SX}^{ij}$ are mass parameters around the GUT scale.
The corresponding mass matrix for the basis 
$(S_{\Phi}, S, S_{\Phi'}, S', N_{\chi}^c, \overline{N}_{\overline{\chi}}^c) $
are
\begin{equation}
\label{MSinglets}
M_{\rm singlets} ~=~{1\over 2} ~\left(
\begin{array}{cccccc}
M_{SX}^{11} & ~~0 & ~~M_{SX}^{12} & ~~0 & ~~0 &  ~~\lambda_3 \langle N_{H}^c \rangle \\
0 &  ~~0 & ~~0 &  ~~0 & ~~0 &  \lambda_3 \lambda_4 \langle N_{H}^c \rangle \\
M_{SX}^{12} &  ~~0 & ~~M_{SX}^{22} &  ~~0 & 
~~\lambda_5 \langle \overline{N}_{\overline H}^c \rangle &  ~~0 \\
0 &  ~~0 & ~~0 &  ~~0 & 
~~\lambda_5 \lambda_6 \langle \overline{N}_{\overline H}^c \rangle &  ~~0 \\
0 &  ~~0 & ~~\lambda_5 \langle \overline{N}_{\overline H}^c \rangle &
~~\lambda_5 \lambda_6 \langle \overline{N}_{\overline H}^c \rangle & ~~0  & ~~M_{\chi} \\
\lambda_3 \langle N_{H}^c \rangle & ~~\lambda_3 \lambda_4 \langle N_{H}^c \rangle
& ~~0 &  ~~0 & ~~M_{\chi} & ~~0 \end{array}
\right)~.~\,
\end{equation}
The determinant of above mass matrix is
\begin{eqnarray}
{\rm Det}[M_{\rm singlets}] &=& {1\over {64}} \lambda_3^2 \lambda_4^2 \lambda_5^2
\lambda_6^2 \left[ M_{SX}^{11} M_{SX}^{22} -
(M_{SX}^{12})^2  \right]  (\langle \overline{N}_{\overline H}^c \rangle)^2
(\langle N_{H}^c \rangle)^2  
\sim M_{GUT}^2 M_{23}^4~.~\,
\end{eqnarray}
Thus, there are two SM singlets (major components from $S_{\Phi}$ and $S_{\Phi'}$)
with masses around the GUT scale, and four SM singlets with masses 
around the scale $M_{23}$. By the way, these SM singlets do not contribute to 
the RGE running below the $M_{23}$ scale.

(3) The SM doublet mixings. After the flipped $SU(5)$ gauge symmetry
breaking, we have the following mass terms in the superpotential for the
SM doublets $H_u$, $H_d$, $L_{\Sigma}$, $\overline{L}_{\overline \Sigma}$,
$L_{\chi}$, and $\overline{L}_{\overline \chi}$
\begin{eqnarray}
W & \supset & -4 \lambda_3 V_{\Phi} L_{\Sigma} \overline{L}_{\overline \chi}
- 4 \lambda_5 V_{\Phi'} L_{\chi} \overline{L}_{\overline \Sigma}
+ 2 \lambda_1 \langle N_{H}^c \rangle L_{\Sigma} H_u
+  2 \lambda_2 \langle \overline{N}_{\overline H}^c  \rangle 
H_d \overline{L}_{\overline \Sigma} 
\nonumber \\ &&
+  \mu H_d H_u
+ M_{\chi} L_{\chi} \overline{L}_{\overline \chi} ~.~\,
\label{WS-Doublets}
\end{eqnarray}
The corresponding mass matrix for the basis 
$(H_d, L_{\Sigma}, L_{\chi})^t$ versus
$(H_u, \overline{L}_{\overline \Sigma}, \overline{L}_{\overline \chi})$ are
the following
\begin{equation}
\label{MDoublets}
M_{doublets} ~=~\left(
\begin{array}{ccc}
 \mu & ~~2 \lambda_2 \langle \overline{N}_{\overline H}^c \rangle & ~~0 \\
2 \lambda_1 \langle N_{H}^c \rangle & ~~0 &
~~ -4 \lambda_3 V_{\Phi} \\
0 & ~~ - 4 \lambda_5 V_{\Phi'} & ~~ M_{\chi}\end{array}
\right)~.~\,
\end{equation}
The determinant of above mass matrix is
\begin{eqnarray}
{\rm Det}[M_{\rm doublets}] &=& - 16 \lambda_3 \lambda_5 \mu V_{\Phi} V_{\Phi'} -
 4 \lambda_1 \lambda_2  M_{\chi} \langle \overline{N}_{\overline H}^c \rangle
\langle N_{H}^c \rangle 
\label{M-doublets}~.~\,
\end{eqnarray}
Note that $V_{\Phi} \sim V_{\Phi'} \sim M_{GUT}$ and 
$\langle \overline{N}_{\overline H}^c \rangle = \langle N_{H}^c \rangle \sim M_{23}$,
we obtain that
there are two pairs of vector-like particles 
(major components belong to $L_{\Sigma}$ and $L_{\chi}$,
and $\overline{L}_{\overline \Sigma}$ and
 $\overline{L}_{\overline \chi}$) with vector-like
masses around the GUT scale, and one pair of 
vector-like particles (major components belong to $H_d$
and $H_u$) whose vector-like mass $M_{LD}$ is
\begin{eqnarray}
M_{LD} & \simeq & {{{\rm Det}[M_{\rm doublets}]}
\over\displaystyle {16 \lambda_3 \lambda_5 V_{\Phi} V_{\Phi'} }} 
\sim  - \mu -{{M_{23}^2}\over {M_{GUT}^2}} M_{\chi}~.~\,
\end{eqnarray}
Because we need one pair of the Higgs doublets with mass around
TeV scale to break the electroweak gauge symmetry,
 we obtain that $\mu$ should be around the TeV scale, and $M_{\chi}$
has a upper bound for a concrete model with gauge coupling unification.
For example, with $M_{23} = 3.66\times 10^{16} ~{\rm GeV}$ and
$M_{GUT} = 4.8 \times 10^{18}~{\rm GeV}$ as in the first
case in the next subsection for gauge coupling unification, 
we obtain that $M_{\chi} \le 1.72\times 10^7 ~{\rm GeV}$.
Moreover, we emphasize that even if $\mu=0$, we can generate the corresponding
effective $\mu_{eff}$ term for one pair of the light Higgs doublets from
above discussions.

With fine-tuning, there are two ways that we can have one pair of
light Higgs doublets and very large vector-like mass $M_{\chi}$ for
$\overline{\chi}_{\overline F}$ and $\chi_F$. One
way is that we fine-tune the two terms in Eq. (\ref{M-doublets})
so that ${\rm Det}[M_{\rm doublets}] \sim \mu_{eff} M_{GUT}^2$ where
$\mu_{eff} \sim 1~{\rm TeV}$. The other way is that
we replace the term $M_{\chi} \overline{\chi} \chi$ in the superpotential
 in Eq. (\ref{SO(10)-WSS}) by the following two terms 
\begin{eqnarray}
W & \supset & y_{\chi}  \overline{\chi} (\Phi - 3 \lambda_4 S) \chi
+ y'_{\chi}  \overline{\chi} (\Phi' - 3 \lambda_6 S') \chi ~,~\,
\label{S-doublets}
\end{eqnarray}
where $y_{\chi}$ and $y_{\chi}'$ are small Yukawa couplings.
Note that
\begin{eqnarray}
\langle \Phi \rangle -3 \lambda_4 \langle S \rangle 
&=& {\rm diag}(4,4,4,0,4,4,4,0,4,4,4,8,0,0,0,4) ~V_{\Phi}~,~\nonumber \\ 
\langle \Phi' \rangle -3  \lambda_6 \langle S' \rangle
&=& {\rm diag} (4,4,4,0,4,4,4,0,4,4,4,8,0,0,0,4) ~V_{\Phi'}~,~\,
\end{eqnarray} 
we have 
\begin{eqnarray}
W & \supset &  4 y_{\chi} V_{\Phi} \left( \overline{\chi}_{\overline F} \chi_F
+ 2 \overline{\chi}_{l} \chi_{\overline l} \right)
+ 4 y_{\chi}' V_{\Phi'} \left( \overline{\chi}_{\overline F} \chi_F
+ 2 \overline{\chi}_{l} \chi_{\overline l} \right)~.~\,
\end{eqnarray} 
Thus, we obtain that the two terms in the superpotential in Eq. (\ref{S-doublets})
will give vector-like masses to $\overline{\chi}_{\overline F}$ and
$\chi_F$, and $\overline{\chi}_{l}$ and $\chi_{\overline l}$, while
they will not give vector-like mass to 
 $\overline{\chi}_{f}$ and $\chi_{\overline f}$. And
then we do not have the last term 
$M_{\chi} L_{\chi} \overline{L}_{\overline \chi}$
in Eq. (\ref{WS-Doublets}), and
the $(3,3)$ entry in the mass matrix in Eq. (\ref{MDoublets})
is zero, {\it i.~e.}, there is no $M_{\chi}$ entry in Eq. (\ref{MDoublets}).
Therefore, the vector-like mass for $\overline{\chi}_{\overline F}$ and
$\chi_F$ can be any value below the $M_{23}$ scale.
By the way, in the concrete model building, we just need one term in 
the superpotential in Eq. (\ref{S-doublets}).

\subsection{Gauge Coupling Unification}

We will study the gauge coupling unification. First, let us
consider the masses for the additional particles.
As discussed in the above subsection, there is one pair of 
vector-like particles (major components belong to $Q_{\chi}$ and 
${\overline Q}_{\overline{\chi}}$) with vector-like mass around $M_{\chi}$.
Also, the particles $D_{\chi}^c$ and $\overline{D}_{\overline{\chi}}^c$
have vector-like mass $M_{\chi}$. For simplicity,
we assume that the correponding vector-like masses for
these particles are the same, and we denote their
masses as $M_V$ because in the fine-tuning case,
we may not have the $M_{\chi} \overline{\chi} \chi$ term in the
superpotential in Eq. (\ref{SO(10)-WSS}).
We also assume that the masses for the color triplets
of $h_{\mathbf{10}}$, $H$, $\overline{H}$, $N_{\chi}^c$, and 
$\overline{N}_{\overline{\chi}}^c$ are around 
the $SU(2)_L\times SU(3)_C$ unification scale $M_{23}$,
and the masses for the $\Sigma_{\bar f}$, $\Sigma_{\bar l}$, 
${\overline \Sigma}_{f}$, ${\overline \Sigma}_{l}$, 
 $\chi_{\bar f}$, $\chi_{\bar l}$, 
${\overline \chi}_{f}$, ${\overline \chi}_{l}$, $\Phi$, and
$\Phi'$ are around the GUT scale $M_{GUT}$, where we do not
write the particles in terms of mass eigenstates here.
Moreover, we denote the $Z$-boson mass as $M_Z$, and
the supersymmetry breaking scale as $M_S$. Also, the order
of mass scales are assumed to be
$M_Z \le M_S \le M_V \le M_{23} \le M_{GUT}$.

For gauge coupling unification, we consider the one-loop 
renormalizaton group equation (RGE) running for the
gauge couplings because the two-loop effects only give
minor corrections as long as the theory is perturbative.
The generic one-loop RGEs for gauge couplings are
\begin{eqnarray}
(4\pi)^2\frac{d}{dt}~ g_i &=& b_i g_i^3~,~\,
\label{SMgauge}
\end{eqnarray}
where $t=\ln  \mu$ with $ \mu$ being the renormalization scale,
$g_1^2 \equiv 5 g_Y^2/3$, and the 
$g_Y$, $g_2$, and $g_3$ are the gauge couplings for
the $U(1)_Y$, $SU(2)_L$, and $SU(3)_C$ gauge groups, respectively.

The gauge coupling unification for the flipped $SU(5)$ is
realized by first unifying $\alpha_2$ and $\alpha_3$ at scale
$M_{23}$, then the gauge couplings of $SU(5)$ and $U(1)_X$ further
unify at the scale $M_{GUT}$.  From $M_{Z}$ to $M_{S}$, 
the beta functions are $b^0 \equiv (b_1, b_2, b_3) = (41/10,-19/6,-7)$, 
and from $M_{S}$ to $M_{V}$, the beta functions are 
$b^I =(33/5,1,-3)$.  From $M_{V}$ to the $\alpha_2$ and $\alpha_3$
unification scale $M_{23}$, the beta functions are $b^{II} =(36/5,4,0)$.

Unification of $\alpha_2$ and $\alpha_3$ at 
the scale $M_{23}$ gives the condition
\begin{eqnarray}
\alpha_2^{-1}(M_Z) - \alpha_3^{-1}(M_Z) = &&
 \frac{b_2^{0} - b_3^{0}}{2
  \pi} \log \left(\frac{M_{SUSY}}{M_Z}\right)
+ \frac{b_2^{I} - b_3^{I}}{2
  \pi} \log \left(\frac{M_{V}}{M_{SUSY}}\right)
\nonumber\\ &&
+\frac{b_2^{II} - b_3^{II}}{2
  \pi} \log \left(\frac{M_{23}}{M_{V}}\right)~,
\label{eq:su5uni1}
\end{eqnarray}
which can be solved to obtain the scale $M_{23}$.

The coupling $\alpha_1'$ of $U(1)_X$ is related to $\alpha_1$ and
$\alpha_5$ at the scale $M_{23}$ by
\be
\alpha_1'^{-1}(M_{23}) = \frac{25}{24} \alpha_1^{-1}(M_{23}) -
\frac{1}{24} \alpha_5^{-1}(M_{23})~.~\,
\ee
And above the scale $M_{23}$, the
beta functions for $U(1)_X$ and $SU(5)$ are 
$b^{III} \equiv (b_1', b_5)= (8, -2)$.


\begin{figure}[htb]
\centering
\includegraphics[width=12cm]{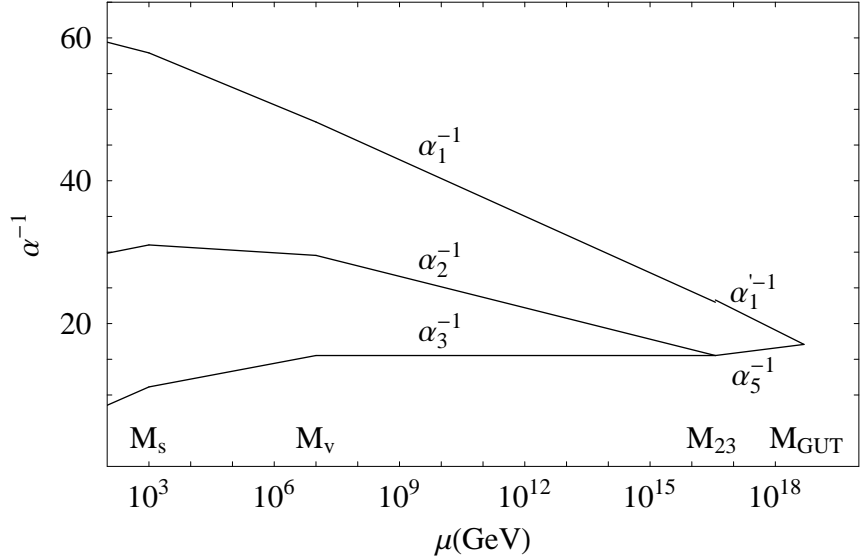}
\caption{The one-loop gauge coupling unification 
for $M_S=10^3$ GeV and $M_V=10^7$ GeV.}
\label{fig:plot1}
\end{figure}


In our numerical calculations, we choose the central values of the strong coupling constant
$\alpha_3(M_Z) = 0.1182 \pm 0.0027$~\cite{Bethke:2004uy}, and the
fine structure constant $\alpha_{EM}$, and
 weak mixing angle $\theta_W$  at $M_Z$ to be~\cite{Eidelman:2004wy}
\bea
\alpha^{-1}_{EM}(M_Z) = 128.91 \pm 0.02~,~~
\sin^2\theta_W(M_Z) = 0.23120 \pm 0.00015\,.~\,
\eea

Because the  top quark pole mass is 
$172.7 \pm 2.9$ GeV~\cite{:2005cc}, 
we might need supersymmetry breaking
scale around or above the TeV scale  
to generate the large enough mass
for the lightest CP even Higgs boson in the MSSM.
So, we assume that $M_S=10^3~{\rm GeV}$.
With $M_V=10^7~{\rm GeV}$, 
we plot the gauge coupling unification in 
Fig.~\ref{fig:plot1}. We obtain that
$M_{23}=3.66\times 10^{16}~{\rm GeV}$, and
$M_{GUT}=4.8\times 10^{18}~{\rm GeV}$.
Note that $M_{23}^2 M_V/M_{GUT} < 10^3~{\rm GeV}$,
we can have one pair of light Higgs doublets without
any fine-tuning. 

Since the GUT scale is close to the Planck scale
 $1.2\times 10^{19}$ GeV, we may need to include
the one-loop supergravity contributions to the RGE running. 
It is reasonable to assume that similar to 
the non-supersymmetric gravity theory~\cite{Robinson:2005fj}, the supergravity
contributions to the one-loop RGEs of gauge couplings are still
proportional to the gauge couplings linearly with 
the same coefficients
for all the gauge couplings because the gravitons and 
gravitinos do not carry any gauge charge.
Note that the gauge coupling of $U(1)_X$ is just a little
bit smaller than that of $SU(5)$ at the
renormalization scale close to the GUT scale,
the supergravity contributions will only slightly increase the 
GUT scale~\cite{Robinson:2005fj}.


\begin{figure}[htb]
\centering
\includegraphics[width=12cm]{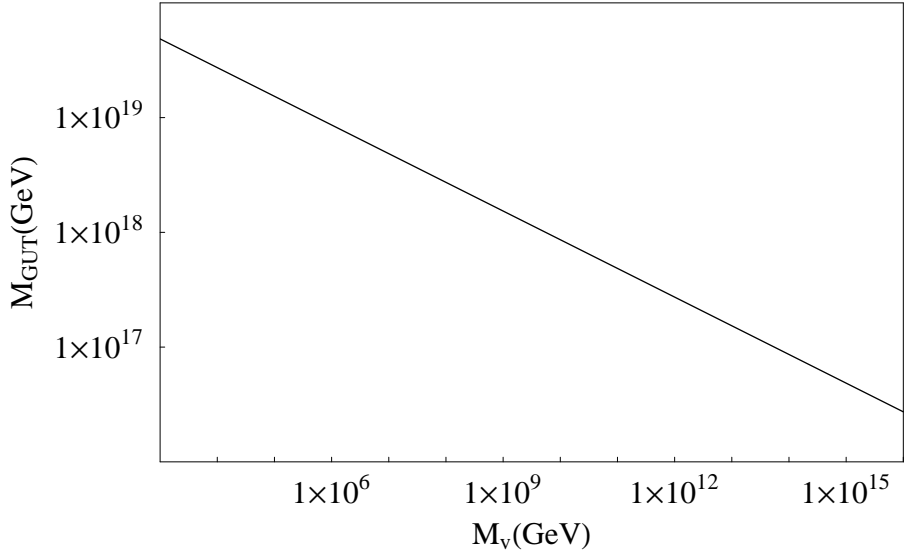}
\caption{ The GUT scale $M_{GUT}$ versus $M_{V}$ for
$M_S=10^3~{\rm GeV}$, and
$M_V$ from $10^3~{\rm GeV}$ to $10^{16}~{\rm GeV}$.}
\label{fig:plot2}
\end{figure}


As discussed in the above subsection, with fine-tuning we can
have very large $M_V$. Assuming $M_S=10^3~{\rm GeV}$,
 we plot the GUT scale $M_{GUT}$ versus $M_{V}$ for
$M_V$ from $10^3~{\rm GeV}$ to $10^{16}~{\rm GeV}$ in
 Fig.~\ref{fig:plot2}. Varying $M_V$ will not
change the scale $M_{23}$ because these vector-like particles
contribute the same one-loop beta functions to
$SU(2)_L$ and $SU(3)_C$. Generically speaking, increasing
$M_V$ will decrease the GUT scale. In addition to the
threshold corrections at the supersymmetry breaking
scale due to the mass differences of the sparticles, 
it is well-known that there
 exist a few percent threshold corrections at the 
GUT scale in the concrete GUT models. So, the gauge coupling
unification for $M_V$ close to $10^6$ GeV is still
fine although there exists less than one percent discrepancy
between the gauge couplings $\alpha_i^{-1}$.
It is interesting to have the GUT scale $M_{GUT}$ 
around the string scale from $10^{17}~{\rm GeV}$
to $10^{18}~{\rm GeV}$, and we find that
 the corresponding $M_V$ scale is from
$5.54\times 10^{13}~{\rm GeV}$ to 
$5.54\times 10^{9}~{\rm GeV}$.

\begin{figure}[htb]
\centering
\includegraphics[width=12cm]{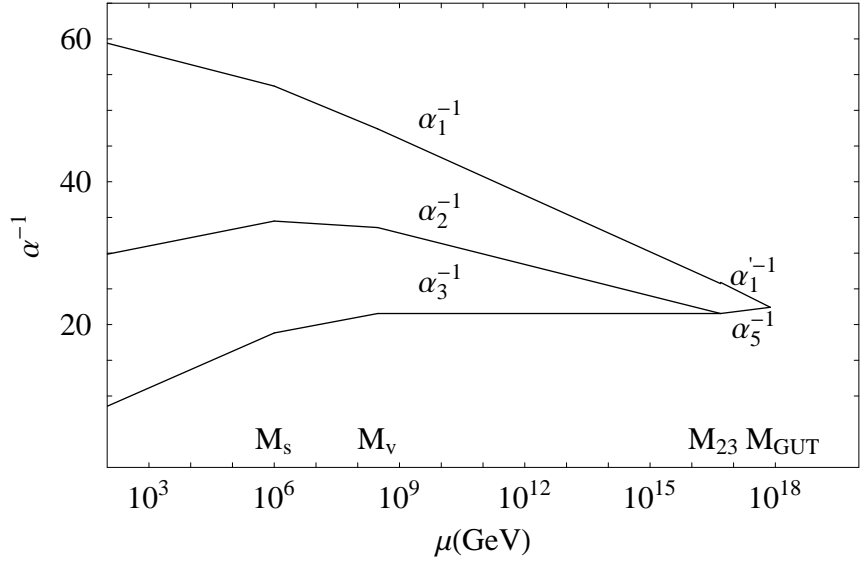}
\caption{The one-loop gauge coupling unification 
for $M_S=10^6$ GeV and $M_V=3\times 10^8$ GeV.}
\label{fig:plot3}
\end{figure}

High-scale supersymmetry breaking~\cite{HSUSY, NASD, Barger:2004sf}
 is interesting due to the appearance of the string landscape~\cite{String} where we may explain 
the cosmological constant problem and gauge hierarchy
 problem~\cite{Weinberg, Agrawal:1998xa}, and  all the problems 
related to the
low energy supersymmetry will be solved automatically
 if the supersymmetry breaking scale is higher than the 
PeV ($10^{15}~{\rm eV}\equiv 10^6~{\rm GeV}$) scale~\cite{Wells:2003tf}.
 Assuming $M_S=10^6~{\rm GeV}$ and $M_V=3\times 10^8~{\rm GeV}$, 
we plot the gauge coupling unification in 
Fig.~\ref{fig:plot3}. We obtain that
$M_{23}=4.88\times 10^{16}~{\rm GeV}$, and
$M_{GUT}=7.57\times 10^{17}~{\rm GeV}$.
Note that $M_{23}^2 M_V/M_{GUT} \sim 1.25 \times 10^6~{\rm GeV}$,
we can also have one pair of light Higgs doublets 
at the PeV scale without fine-tuning. By the way,
the SM Higgs doublet with electroweak-scale mass
is obtained by fine-tuning the mass matrix
for the scalar Higgs doublets.


\section{Remarks}

We would like to briefly discuss the simplest 
$SO(10)$ model with flipped $SU(5)$
embedding where there is only one adjoint Higgs field,
and we point out its major phenomenological
difficulty. We will also  explain how to
generate the small mass for $M_{\chi}$.

\subsection{$SO(10)$ Model with One Adjoint Higgs Field}

We can embed the flipped $SU(5)$ models
into the $SO(10)$ model with only one adjoint Higgs field $\Phi$.
In the superpotential in Eq. (\ref{SO(10)-WSS}),
we change $W(\Phi, \Phi')$ to $W(\Phi)$, and replace 
the $\lambda_5 \overline{\Sigma} (\Phi' + \lambda_6 S') \chi$
term  by the following term
\begin{eqnarray}
W & \supset & \lambda_5 \overline{\Sigma}
 (\Phi + \lambda_6 S') \chi~.~
\end{eqnarray}
The discussions for the splittings of the five/one-plets and ten-plets 
in the multiplets $\overline{\chi}$ and $\Sigma$, and 
 $\overline{\Sigma}$ and $\chi$,
 are the same as those in the Section III
except that we replace $\Phi'$ by $\Phi$, and $V_{\Phi'}$ by
$V_{\Phi}$.

Let us concentrate on the problem. The mass matrix for the basis 
$({\overline Q}_{\Phi}, {\overline Q}_{\overline{\chi}})^t$ versus
$(Q_{\Phi},  Q_{\chi})$,  are the following
\begin{equation}
\label{Remarks-MXYQ}
M_{XYQ\overline{Q}} ~=~\left(
\begin{array}{ccc}
M_{XY}^{11} &  ~~\lambda_5 \langle \overline{N}_{\overline H}^c \rangle \\
\lambda_3 \langle N_{H}^c \rangle  & ~~M_{\chi} \end{array}
\right)~.~\,
\end{equation}
The determinant of above mass matrix is
\begin{eqnarray}
{\rm Det}[M_{XYQ\overline{Q}}] &=& M_{XY}^{11} M_{\chi} 
- \lambda_3 \lambda_5 \langle \overline{N}_{\overline H}^c \rangle
\langle N_{H}^c \rangle~.~\,
\end{eqnarray}
The discussions for the mass matrix of SM doublets are the same
as those in the subsection A in Section IV 
except that we change $V_{\Phi'}$
to $V_{\Phi}$ in Eqs. (\ref{WS-Doublets}) and (\ref{MDoublets}). 
So, without fine-tuning 
the $M_{\chi}$  still cannot be larger than about $10^{8}~{\rm GeV}$.
Then we have 
\begin{eqnarray}
{\rm Det}[M_{XYQ\overline{Q}}] & \sim &
- \lambda_3 \lambda_5 \langle \overline{N}_{\overline H}^c \rangle
\langle N_{H}^c \rangle  \sim - M_{23}^2~.~\,
\end{eqnarray}
Thus,  there is one pair of vector-like particles 
(major components belong to $Q_{\Phi}$  and
 ${\overline Q}_{\Phi}$) with vector-like 
mass around the GUT scale, and one pair of 
vector-like particles (major components belong to $Q_{\chi}$ and 
${\overline Q}_{\overline{\chi}}$) with vector-like mass around 
$M_{23}^2/M_{GUT}$.
Note that the particles $D_{\chi}^c$ and $\overline{D}_{\overline{\chi}}^c$
have vector-like mass $M_{\chi}$,
 we can easily show that the gauge coupling 
unification can not be realized. By the way,
with large fine-tuning so that $M_{\chi}$ can be 
around $M_{23}^2/M_{GUT}$ and 
${\rm Det}[M_{XYQ\overline{Q}}] \sim 10^{-2} M_{23}^2$,
we can have gauge coupling unification.

With $M_{\chi} \le 10^{8}~{\rm GeV}$ and without fine-tuning, 
we may also achieve the gauge coupling 
unification by adding extra vector-like particles, for
example, one or two pairs of $\mathbf{\overline{16}}$ and
$\mathbf{16}$. However, these models are very complicated
in general, and still need some fine-tuning to achieve
the gauge coupling unification after detailed study.

\subsection{Explanation to the Suitable Mass $M_{\chi}$}

To have the natural models, we need to explain why $M_{\chi}$ can
be around $10^{7}~{\rm GeV}$. There are two well-known ways
to generate  small masses: the Froggat-Nielsen mechanism~\cite{FN} 
and the see-saw mechanism~\cite{Seesaw}. Because we will try to generate the
SM fermion masses and mixings, and the suitable mass $M_{\chi}$ via 
Froggat-Nielsen mechanism by introducing extra flavour 
symmetry in our models in a future publication,
 we  employ the see-saw mechanism to explain the $M_{\chi}$ here.

As we know, an elegant and popular solution to the strong CP problem
is  the Peccei-Quinn mechanism~\cite{PQ}, in which a global axial 
symmetry $U(1)_{PQ}$ is introduced and broken spontaneously at 
some high energy scale. The original Weinberg--Wilczek axion~\cite{WW} is 
excluded by experiment, in particular by the non-observation of 
the rare decay $K \rightarrow \pi +a$~\cite{review} where
$a$ is the axion field.  There are two viable ``invisible'' axion models 
in which the experimental bounds can be evaded: 
(1)~the Kim--Shifman--Vainshtein--Zakharov  (KSVZ) axion model, 
which introduces a SM singlet $S_{PQ}$ and a pair of extra 
vector-like quarks that carry $U(1)_{PQ}$ charges while the SM fermions 
and Higgs fields are neutral under $U(1)_{PQ}$ symmetry~\cite{KSVZ}; 
(2)~the 
Dine--Fischler--Srednicki--Zhitnitskii (DFSZ) axion model, in which a 
SM singlet $S_{PQ}$ and one pair of Higgs doublets are introduced,
and the SM fermions and Higgs fields 
are also charged under $U(1)_{PQ}$ symmetry~\cite{DFSZ}.
From laboratory, astrophysical, and cosmological constraints, the $U(1)_{PQ}$
symmetry breaking scale  is limited to  the range from $10^{10}~{\rm GeV}$
to  $10^{12}~{\rm GeV}$~\cite{review}.
And then the VEV of $S_{PQ}$ is also roughly in the range
from $10^{10}~{\rm GeV}$ to  $10^{12}~{\rm GeV}$. Interestingly, 
$(\langle S_{PQ} \rangle)^2/M_{23} $ can be from $10^{4}~{\rm GeV}$
to $10^{8}~{\rm GeV}$, which can give us  the needed
mass scale for $M_{\chi}$.

Let us introduce one pair of the spinor $\mathbf{\overline{16}}$ and
$\mathbf{16}$ representation 
vector-like particles $\overline{\chi}'$ and $\chi'$.
In the superpotential in Eq. (\ref{SO(10)-WSS}),
we can forbid the $M_{\chi} \overline{\chi} \chi$
term by $U(1)_{PQ}$ symmetry, and introduce the following superpotential
\begin{eqnarray}
W & \supset & M_{\chi'} \overline{\chi}' \chi' +
\lambda_{PQ1} S_{PQ} \overline{\chi}' \chi + 
\lambda_{PQ2} S_{PQ} \overline{\chi} \chi' ~,~\,
\label{Com-Chi-SS}
\end{eqnarray}
where $\lambda_{PQ1}$ and $\lambda_{PQ2}$ are the Yukawa couplings,
and $M_{\chi'}$ is a mass parameter around the scale $M_{23}$
which can be generated via Froggat-Nielsen mechanism easily.

Because we are not interested in the superheavy states that
are always superheavy without fine-tuning, 
let us focus on the mixings between the light states 
$\overline{\chi}_{\overline F}$  and $\chi_F$  of
 $\overline{\chi}$ and $\chi$ and the superheavy states
$\overline{\chi}'_{\overline F}$  and $\chi'_F$  of
 $\overline{\chi}'$ and $\chi'$.
After the $U(1)_{PQ}$ symmetry breaking,  
the mass matrix for the basis 
$(\overline{\chi}_{\overline F}, \overline{\chi}'_{\overline F})^t$ versus
$(\chi_F, \chi'_F)$ is
\begin{equation}
\label{Mchichip}
M_{\chi_F\chi'_F} ~=~\left(
\begin{array}{cc}
0 & ~~ \lambda_{PQ2} \langle S_{PQ} \rangle \\
\lambda_{PQ1} \langle S_{PQ} \rangle &~~ M_{\chi'} \end{array}
\right)~.~\,
\end{equation}
Thus, we obtain that there is one pair of vector-like particles 
(major components belong to $\overline{\chi}'_{\overline F}$ 
and $\chi'_{F}$) with vector-like 
mass around the GUT scale, and one pair of 
vector-like particles (major components belong to $\chi_F$ and 
$\overline{\chi}_{\overline F}$) with vector-like mass around 
\begin{eqnarray}
M_{\rm light ~\chi_F} \sim {{\lambda_{PQ1} \lambda_{PQ2} 
(\langle S_{PQ} \rangle)^2}\over\displaystyle {M_{\chi'}}}
\sim 10^{4-8}~{\rm GeV}~.~\
\label{Com-Chi-ML}
\end{eqnarray}

In fact, we can simply integrate out vector-like particles 
$\overline{\chi}'$ and $\chi'$ in Eq. (\ref{Com-Chi-SS}),
and obtain the following superpotential
\begin{eqnarray}
W & \supset &-
\lambda_{PQ1} \lambda_{PQ2} 
{{S^2_{PQ}}\over {M_{\chi'}}} \overline{\chi} \chi  ~.~\,
\end{eqnarray}
This is the exact high-dimensional operator that
can generate the suitable vector-like mass  $M_{\chi}$.
In short, we can indeed generate the light $M_{\chi}$ 
naturally.



\section{Discussion and Conclusions}

We embedded the flipped $SU(5)$ models into the 
$SO(10)$ models. After  the $SO(10)$ gauge symmetry is broken 
down to the flipped $SU(5)$ gauge symmetry,  we can 
 split the five/one-plets and ten-plets in the multiplets
$\overline{\chi}$ and $\Sigma$, and  $\overline{\Sigma}$ and $\chi$ via 
 the stable sliding singlet mechanism. Similar to the
flipped $SU(5)$ model, the gauge symmetry can be
broken down to the SM gauge symmetry by giving VEVs to
the singlet componets of $H$ and $\overline{H}$.
The doublet-triplet splitting problem can be solved  
naturally by the missing partner mechanism, and 
the Higgsino-exchange mediated proton decay can be
avoided elegantly. Moreover,
we showed that there exists one pair of the light Higgs doublets 
with major components from $H_u$ and $H_d$ for the
electroweak gauge symmetry breaking.
Because there exist two pairs of the vector-like fields 
with similar intermediate-scale masses
(major components from $Q_{\chi}$ and 
${\overline Q}_{\overline{\chi}}$, and 
$D_{\chi}^c$ and $\overline{D}_{\overline{\chi}}^c$),
we can have gauge coupling unification at the GUT scale which is 
 reasonably (about one or two orders) higher than 
the $SU(2)_L\times SU(3)_C$ unification scale.
In short, we can keep the beautiful features  and get rid of the drawbacks 
of the flipped $SU(5)$ models in our $SO(10)$ models.

Furthermore, we briefly studied the simplest $SO(10)$ model
with flipped $SU(5)$ embedding, and found that
it can not work without fine-tuning.
We also explained how to generate the suitable vector-like mass 
$M_{\chi}$ for $\chi$ and $\overline{\chi}$.

\begin{acknowledgments}

T.L. would like to thank S.~M.~Barr and J.~Jiang for helpful discussions.
The research of T.L. was supported by DOE grant DE-FG02-96ER40959.

\end{acknowledgments}

\appendix

\section{ The $SO(10)$ Generators in the Spinor  Representations}
\label{apdxA}

The $SO(10)$  generators in the spinor  representations 
and the assignment of the SM fermions in the ${\mathbf{16}}$ can
be found in Ref.~\cite{Rajpoot:xy}.  We copy the $\sigma \cdot
W_\mu$, and rename it as $/\!\!\!\!{A_\mu}$.  The $16\times 16$ matrix
for $/\!\!\!\!{A_\mu}$ can be re-written into 
the following four $8\times 8$ matrices
\be
\label{eq:gauge16}
/\!\!\!\!{A} =
\left(
\begin{array}{cc}
/\!\!\!\!{A_{11}}&/\!\!\!\!{A_{12}} \\
/\!\!\!\!{A_{21}}&/\!\!\!\!{A_{22}}
\end{array}
\right)\,,
\ee
with
\bea
/\!\!\!\!{A_{11}} &=&
\left(
\begin{array}{cccccccc}
\ld_{1 1}&V_{12}&V_{13}&X_1^0&W_L^{-}& & & \\
V_{12}^*&\ld_{2 2}&V_{23}&X_2^{-}& &W_L^{-}& & \\
V_{13}^*&V_{23}^*&\ld_{3 3}&X_3^{-}& & &W_L^{-}& \\
\overline{X}_1^0&X_2^+&X_3^+&\ld_{4 4}& & &
&W_L^{-}\\
W_L^+& & & &\ld_{5 5}&V_{12}&V_{13}&X_1^0 \\
 &W_L^+& & &V_{12}^*&\ld_{6 6}&V_{23}&X_2^{-} \\
 & &W_L^+& &V_{13}^*&V_{23}^*&\ld_{7 7}&X_3^{-} \\
 & & &W_L^+&\overline{X}_1^0&X_2^+&X_3^+&\ld_{8 8}
\end{array}
\right)\,, \nnb \\
/\!\!\!\!{A_{12}} &=& \left(
\begin{array}{cccccccc}
0&A_6^0&-A_5^0&-Y_1^+&0&-Y_6^{-}&Y_5^{-}&-\overline{A}_1^0 \\
-A_6^0&0&A_4^{-}&-\overline{Y}_2^0&Y_6^{-}&0&-Y_4^{--}&-A_2^{-} \\
A_5^0&-A_4^{-}&0&-\overline{Y}_3^0&-Y_5^{-}&Y_4^{--}&0&-A_3^{-} \\
Y_1^+&\overline{Y}_2^0&\overline{Y}_3^0&0&\overline{A}_1^0
&A_2^{-}&A_3^{-}&0\\
0&-A_3^+&A_2^+&-Y_4^{++}&0&Y_3^0&-Y_2^0&-A_4^+ \\
A_3^+&0&A_1^0&-Y_5^+&Y_3^0&0&-Y_1^{-}&-\overline{A}_5^0 \\
-A_2^+&A_1^0&0&-Y_6^+&Y_2^0&-Y_1^{-}&0&-\overline{A}_6^0 \\
Y_4^{++}&Y_5^+&Y_6^+&0&A_4^+&\overline{A}_5^0&\overline{A}_6^0&0
\end{array}
\right)\,, \nnb \\
/\!\!\!\!{A_{21}} &=& \left(
\begin{array}{cccccccc}
0&-\overline{A}_6^0&\overline{A}_5^0&Y_1^{-}&0&A_3^{-}&-A_2^{-}&Y_4^{--}\\
\overline{A}_6^0&0&-A_4^+&Y_2^0&-A_3^{-}&0&\overline{A}_1^0&Y_5^{-} \\
-\overline{A}_5^0&A_4^+&0&Y_3^0&A_2^{-}&-\overline{A}_1^0&0&Y_6^{-} \\
-Y_1^{-}&-Y_2^0&-Y_3^0&0&-Y_4^{-}&-Y_5^{-}&-Y_6^{-}&0 \\
0&Y_6^+&-Y_5^+&A_1^0&0&-\overline{Y}_3^0&\overline{Y}_2^0&A_4^{-} \\
-Y_6^+&0&Y_4^{++}&A_2^+&\overline{Y}_3^0&0&-Y_1^+&A_5^0 \\
Y_5^+&-Y_4^{++}&0&A_3^+&-\overline{Y}_2^0&Y_1^+&0&A_6^0 \\
-A_1^0&-A_2^+&-A_3^+&0&-A_4^{-}&-A_5^0&-A_6^0&0
\end{array}
\right)\,, \nnb \\
/\!\!\!\!{A_{22}} &=& \left(
\begin{array}{cccccccc}
\ld_{9 9}&-V_{12}^*&-V_{13}^*&-\overline{X}_1^0&W_R^{-}& & & \\
-V_{12}&\ld_{10 10}&-V_{23}^*&-X_2^+& &W_R^{-}& & \\
-V_{13}&-V_{23}&\ld_{11 11}&-X_3^+& & &W_R^{-}& \\
-X_1^0&-X_2^{-}&-X_3^{-}&\ld_{12 12}& & & &W_R^{-} \\
W_R^+& & & &\ld_{13 13}&-V_{12}^*&-V_{13}^*&-\overline{X}_1^0 \\
 &W_R^+& & &-V_{12}&\ld_{14 14}&-V_{23}^*&-X_2^+ \\
 & &W_R^+& &-V_{13}&-V_{23}&\ld_{15 15}&-X_3^+ \\
 & & &W_R^+&-X_1^0&-X_2^-&-X_3^3&\ld_{16 16}
\end{array}
\right)~.~ \nnb
\eea

The 45 gauge bosons consist of 12 $A$, 6 $X$, 6 $V$, 12 $Y$, 2
charged $W_L$, 2 charged $W_R$, and 16 $\lambda$ which can be rewritten
as 5 independent fields, $V_3$, $V_8$, $V_{15}$, $W_L^0$ and $W_R^0$.

The first family of the SM fermions forms a 
spinor ${\mathbf{16}}$ representation
\be 
{\mathbf{16}}_1=
(u_r,u_g,u_b,\nu_e,d_r,d_g,d_b,e^-,d_r^c,d_g^c,d_b^c,e^+,-u_r^c,-u_g^c,-u_b^c,-\nu_e^c)^t~,
\ee 
similarly for the second and third families.  As the $SO(10)$ is
broken down to $SU(5) \times U(1)$ or flipped $SU(5)$,
the spinor representation ${\mathbf{16}}$ is decomposed as
\bea
{\mathbf {16}} \to
({\mathbf{ 10, 1}}) + ({\mathbf{ {\overline 5}, -3}}) + ({\mathbf{ 1, 5}}) ~,~\,
\eea
where 
\be ({\mathbf{ 10, 1}}) =
(Q,U^c,E^c), \quad ({\mathbf{ {\overline 5}, -3}}) = (D^c,L), \quad {\rm and} \quad 
({\mathbf{ 1, 5}}) = N^c 
\ee 
for breaking to $SU(5) \times U(1)$, and 
\be 
({\mathbf{ 10, 1}})
= (Q,D^c,N^c), \quad ({\mathbf{ {\overline 5}, -3}}) = (U^c,L), \quad {\rm and} \quad
({\mathbf{ 1, 5}}) = E^c
\ee 
for breaking to flipped $SU(5)$.

\end{document}